\documentstyle[12pt]{article}

\def\hybrid{\topmargin -20pt    \oddsidemargin 0pt
        \headheight 0pt \headsep 0pt
        \textwidth 6.25in       
        \textheight 9.5in       
        \marginparwidth .875in
        \parskip 5pt plus 1pt   \jot = 1.5ex}

\hybrid

\def\baselinestretch{1.2}

\catcode`\@=11

\def\marginnote#1{}
\newcount\hour
\newcount\minute
\newtoks\amorpm
\hour=\time\divide\hour by60
\minute=\time{\multiply\hour by60 \global\advance\minute by-\hour}
\edef\standardtime{{\ifnum\hour<12 \global\amorpm={am}%
        \else\global\amorpm={pm}\advance\hour by-12 \fi
        \ifnum\hour=0 \hour=12 \fi
        \number\hour:\ifnum\minute<10 0\fi\number\minute\the\amorpm}}
\edef\militarytime{\number\hour:\ifnum\minute<10 0\fi\number\minute}

\def\draftlabel#1{{\@bsphack\if@filesw {\let\thepage\relax
   \xdef\@gtempa{\write\@auxout{\string
      \newlabel{#1}{{\@currentlabel}{\thepage}}}}}\@gtempa
   \if@nobreak \ifvmode\nobreak\fi\fi\fi\@esphack}
        \gdef\@eqnlabel{#1}}
\def\@eqnlabel{}
\def\@vacuum{}
\def\draftmarginnote#1{\marginpar{\raggedright\scriptsize\tt#1}}

\def\draft{\oddsidemargin -.5truein
        \def\@oddfoot{\sl preliminary draft \hfil
        \rm\thepage\hfil\sl\today\quad\militarytime}
        \let\@evenfoot\@oddfoot \overfullrule 3pt
        \let\label=\draftlabel
        \let\marginnote=\draftmarginnote
   \def\@eqnnum{(\theequation)\rlap{\kern\marginparsep\tt\@eqnlabel}%
\global\let\@eqnlabel\@vacuum}  }

\def\preprint{\twocolumn\sloppy\flushbottom\parindent 2em
        \leftmargini 2em\leftmarginv .5em\leftmarginvi .5em
        \oddsidemargin -.5in    \evensidemargin -.5in
        \columnsep .4in \footheight 0pt
        \textwidth 10.in        \topmargin  -.4in
        \headheight 12pt \topskip .4in
        \textheight 6.9in \footskip 0pt
        \def\@oddhead{\thepage\hfil\addtocounter{page}{1}\thepage}
        \let\@evenhead\@oddhead \def\@oddfoot{} \def\@evenfoot{} }

\def\numberbysection{\@addtoreset{equation}{section}
        \def\theequation{\thesection.\arabic{equation}}}

\def\underline#1{\relax\ifmmode\@@underline#1\else
        $\@@underline{\hbox{#1}}$\relax\fi}

\def\titlepage{\@restonecolfalse\if@twocolumn\@restonecoltrue\onecolumn
     \else \newpage \fi \thispagestyle{empty}\c@page\z@
        \def\thefootnote{\fnsymbol{footnote}} }

\def\endtitlepage{\if@restonecol\twocolumn \else \newpage \fi
        \def\thefootnote{\arabic{footnote}}
        \setcounter{footnote}{0}}  

\catcode`@=12
\relax

\def\figcap{\section*{Figure Captions\markboth
        {FIGURECAPTIONS}{FIGURECAPTIONS}}\list
        {Figure \arabic{enumi}:\hfill}{\settowidth\labelwidth{Figure
999:}
        \leftmargin\labelwidth
        \advance\leftmargin\labelsep\usecounter{enumi}}}
 \relax
\def\tablecap{\section*{Table Captions\markboth
        {TABLECAPTIONS}{TABLECAPTIONS}}\list
        {Table \arabic{enumi}:\hfill}{\settowidth\labelwidth{Table
999:}
        \leftmargin\labelwidth
        \advance\leftmargin\labelsep\usecounter{enumi}}}
 \relax
\def\reflist{\section*{References\markboth
        {REFLIST}{REFLIST}}\list
        {[\arabic{enumi}]\hfill}{\settowidth\labelwidth{[999]}
        \leftmargin\labelwidth
        \advance\leftmargin\labelsep\usecounter{enumi}}}
 \relax
\makeatletter
\newcounter{pubctr}
\def\publist{\@ifnextchar[{\@publist}{\@@publist}}
\def\@publist[#1]{\list
        {[\arabic{pubctr}]\hfill}{\settowidth\labelwidth{[999]}
        \leftmargin\labelwidth
        \advance\leftmargin\labelsep
        \@nmbrlisttrue\def\@listctr{pubctr}
        \setcounter{pubctr}{#1}\addtocounter{pubctr}{-1}}}
\def\@@publist{\list
        {[\arabic{pubctr}]\hfill}{\settowidth\labelwidth{[999]}
        \leftmargin\labelwidth
        \advance\leftmargin\labelsep
        \@nmbrlisttrue\def\@listctr{pubctr}}}
 \relax
\makeatother
\newskip\humongous \humongous=0pt plus 1000pt minus 1000pt

\newif\ifdtup

\relax

\def\be{\begin{equation}}
\def\ee{\end{equation}}
\def\ba{\begin{eqnarray}}
\def\ea{\end{eqnarray}}

\def\del{\partial}

\def\r{\rho}
\def\a{\alpha}

\def\b{\beta}

\def\d{\delta}

\def\e{\epsilon}

\def\th{\theta}

\def\m{\mu}
\def\n{\nu}

\def\l{\lambda}
\def\L{\Lambda}
\def\s{\sigma}

\def\cL{{\cal L}}

\def\bs{\bigskip}
\def\no{\noindent}

\def\qq{\qquad}

\def\IR{\relax{\rm I\kern-.18em R}}

\def \z { {\bar z} }

\def \tA  { {\tilde {A }}}

\def \ha {{1\over 2}}

\def \ov {\over}

\def\diag{{\rm diag}}

\def\IR{\relax{\rm I\kern-.18em R}}
\def\inv{^{\raise.15ex\hbox{${\scriptscriptstyle -}$}\kern-.05em 1}}

\def\PL{Poisson--Lie T-duality}

\def\cL{{\cal L}}

\def\tX{{\tilde X}}

\def\tL{{\tilde L}}

\begin{document}

\renewcommand{\theequation}{\arabic{equation}}

\newcommand{\beq}{\begin{equation}}
\newcommand{\eeq}[1]{\label{#1}\end{equation}}
\newcommand{\ber}{\begin{eqnarray}}
\newcommand{\eer}[1]{\label{#1}\end{eqnarray}}
\newcommand{\eqn}[1]{(\ref{#1})}
\begin{titlepage}
\begin{center}

\hfill CERN-TH/98-44\\
\hfill hep--th/9803019\\

\vskip .8in

{\large \bf \PL\ beyond the classical level }
{\large \bf and the renormalization group}

\vskip 0.6in

{\bf Konstadinos Sfetsos}
\vskip 0.1in
{\em Theory Division, CERN\\
     CH-1211 Geneva 23, Switzerland\\
{\tt sfetsos@mail.cern.ch}}\\
\vskip .2in

\end{center}

\vskip .6in

\centerline{\bf Abstract }

\vskip 0,2cm
\no
In order to study quantum aspects of
$\s$-models related by \PL, we construct 
three- and two-dimensional models 
that correspond, in one of the dual faces, to
deformations of $S^3$ and $S^2$.
Their classical canonical 
equivalence is demonstrated by means of a generating functional, which we 
explicitly compute.
We examine how they behave under the renormalization group and show that 
dually related models have the same 1-loop beta functions for 
the coupling and deformation parameters.
We find non-trivial
fixed points in the ultraviolet, where the theories do not become 
asymptotically free. This suggests that the limit 
of \PL\ to the usual Abelian and non-Abelian T-dualities does not exist 
quantum mechanically, although it does so classically.

\vskip 4cm
\noindent
CERN-TH/98-44\\
February 1998\\
\end{titlepage}
\vfill
\eject

\def\baselinestretch{1.2}
\baselineskip 16 pt
\noindent

\def\tT{{\tilde T}}
\def\tg{{\tilde g}}
\def\tL{{\tilde L}}


\section{Introduction}

A generalization of Abelian \cite{BUSCHER} 
and non-Abelian \cite{nonabel} 
target space duality (T-duality) is the so-called \PL\ \cite{KliSevI}. Its 
notable features for our purposes are that it does not 
rely on the existence of isometries but rather on a 
rigid group-theoretical structure \cite{KliSevI} and also that it can 
be explicitly formulated as a canonical transformation between phase-space
variables \cite{PLsfe1,PLsfe2}, 
similarly to ordinary T-duality cases \cite{AALcan,zacloz}.\footnote{\PL\
is by construction a canonical transformation and an abstract expression 
for the generating functional was already given in \cite{klisevCan}. Also,
\PL\ does not cover 
all known T-duality transformations. It does not cover, in particular, 
non-Abelian
duality transformations with respect to vector subgroups, i.e. in WZW models
\cite{gausfe}, 
as well as the vector-axial T-duality \cite{axialvector}.
For this type of dualities a new 
unified framework is required, as explained in \cite{PLsfe2}.}

Up to now all considerations concerning \PL\ were, almost exclusively,
classical. It is the main
objective of the present work to go beyond that level.
The main obstacle in doing so is the (almost complete) lack of explicit 
examples in the
literature of $\s$-models related by genuine \PL. Namely, models 
such that they cannot
be obtained in the context of the usual Abelian or non-Abelian T-duality
and also 
such that their target spaces have the same dimension.
We construct explicit examples in three and two target space dimensions; they 
represent a two- and a one-parameter deformations of 
$S^3$ and $S^2$ and of their non-Abelian duals, respectively, 
and they reduce to them in an appropriate limit. We also give the explicit 
expression for the generating functional of the 
canonical transformation relating them.
We then study renormalization of $\s$-models 
with two-dimensional target spaces by treating them as ordinary 
2-dimensional field theories.
Renormalizability in this context means that
the counter-terms arising at every order in perturbation theory can be absorbed
into renormalizations of the coupling and the parameters, up to field 
redefinitions or, equivalently, diffeomorphisms in the target space.
We show that the 1-loop beta functions for the coupling and the parameters
of the dually related models are equivalent, 
which strongly hints towards their
equivalence beyond the classical level.\footnote{In this respect we note that 
the existence of a canonical transformation relating two 
different $\s$-models seems to be
necessary for their equivalence at the quantum 
level. We note the case of the Principal Chiral model and the 
Pseudodual Chiral model \cite{pscm} whose classical solutions are in one-to-one
correspondence but which are not canonically equivalent \cite{zacloz}. 
It is well known
that their quantum behaviours are drastically different \cite{Nappi}.}
Moreover, we find that their
behaviour in the ultraviolet is completely different
from the corresponding one for $S^2$ and its non-Abelian dual. 
The latter models
are asymptotically free whereas our models have a non-trivial behaviour, i.e.
the geometry is still curved.
This suggests that \PL\ is really distinct from the usual Abelian and 
non-Abelian T-dualities and that it does not correspond to a ``soft breaking''
of these. 
We have also considered renormalization at the 2-loop level. We have
found that the models are
not renormalizable in the strict field-theoretical sense, namely, the 
corresponding counter-term cannot be absorbed into coupling, parameter and 
field renormalization. This presumably implies that the
\PL\ transformation rules should be modified at this level. 

In the rest
of this section we will briefly review,
by following the conventions of \cite{PLsfe1,PLsfe2},
some facts about \PL, which are necessary for our constructions.
The form of 2-dimensional $\s$-model actions related by \PL\
is \cite{KliSevI}\footnote{The most 
general such actions that include spectator extra fields
can be found in \cite{KliSevI,Tyurin}.}
\be
S= {1\ov 2 \l} \int E_{ab} L^a_\m L^b_\n \del_+ X^\m \del_- X^\n~ , 
~~~~ E= (E_0\inv + \Pi)\inv~ ,
\label{action1}
\ee
and 
\be
\tilde S= {1\ov 2 \l} \int \tilde E^{ab} \tL_{a\m} \tL_{b\n} \del_+
 \tilde X^\m \del_- \tilde X^\n~ , 
~~~~ \tilde E=(E_0 + \tilde \Pi)\inv~ .
\label{action2}
\ee
The field variables in \eqn{action1} are
$X^\m$, $\m=1,2,\dots  ,\dim(G)$ and parametrize an element
$g$ of a group $G$.
We also introduce representation matrices
$\{T_a\}$, with $a=1,2,\dots, \dim(G)$ and
the components of the left-invariant Maurer--Cartan forms $L^a_\m$.
The light-cone coordinates on the world-sheet are
$\s^\pm =\ha (\tau \pm \s)$ and $\l$ denotes the overall coupling constant. 
Similarly, for \eqn{action2} the 
field variables are $\tilde X^\m$, where
$\tilde X^\m$, $\m=1,2,\dots  ,\dim(G)$, parametrize a different group 
$\tilde G$, whose dimension is, however, equal to that of $G$.
Accordingly, we introduce a different set 
of representation matrices $\{\tilde T^a\}$, with $a=1,2,\dots, \dim(G)$, and
the corresponding components of the left-invariant Maurer--Cartan forms 
$\tilde  L_{a\m}$. In \eqn{action1} and \eqn{action2},
$E_0$ is a constant $\dim(G) \times \dim(G)$ matrix, whereas 
$\Pi$ and $\tilde \Pi$ are antisymmetric matrices with the 
same dimension as $E_0$, but they depend on the variables $X^\m$ and $\tilde
X^\m$ via the corresponding group elements $g$ and $\tilde g$.
Hence, we do not require any isometry associated 
with the groups $G$ and $\tilde G$. 

It is crucial \cite{KliSevI} for \PL\ that 
the algebras generated by $\{T_a\}$ and $\{\tilde T^a\}$
form a pair 
of maximally isotropic subalgebras into which the Lie algebra of a 
Lie group known as the Drinfeld double can be decomposed.
This implies the existence of non-trivial mixed commutators between 
the $T_a$'s and the $\tilde T^a$'s, of the type we will encounter in section 2.
It is important that there exist also a bilinear invariant 
$\langle{\cdot|\cdot \rangle}$ with the various generators obeying 
\be
\langle{T_a|T_b\rangle}= \langle{\tilde T^a|\tilde T^b \rangle}= 0~ ,
~~~~ \langle{T_a|\tilde T^b \rangle} = \d_a{}^b ~ .
\label{bili}
\ee
\no
The matrices $\Pi$, $\tilde \Pi$ in \eqn{action1} and \eqn{action2}
are defined as
\be
\Pi^{ab} = b^{ca} a_c{}^b ~ , ~~~~~~ 
\tilde \Pi_{ab} = \tilde b_{ca} \tilde a^c{}_b ~ ,
\label{pipi}
\ee
where the matrices $a(g)$, $b(g)$ are constructed using
\be
g\inv T_a g = a_a{}^b T_b~ ,~~~~ g\inv \tilde T^a g = 
b^{ab} T_b +  (a\inv)_b{}^a \tilde T^b~ ,
\label{abpi}
\ee
and similarly for $\tilde a(\tilde g)$ and $\tilde b(\tilde g)$.
Consistency restricts them to obey
\be
a(g\inv) = a\inv(g)~ ,~~~~ b^T(g)= b(g\inv)~ ,~~~~
\Pi^T(g) = - \Pi(g) ~ ,
\label{conss}
\ee 
and similarly for the tilded ones. 

\section{The models }

\subsection{Generalities: algebraic and group-theoretical structure}

We start this chapter by first constructing an explicit Drinfeld double
and a bilinear invariant for it,
and then computing the matrices $a$, $b$, $\Pi$ as well as their tilded 
counterparts.

Consider the 3-dimensional algebras $su(2)$ and $e_3$, with generators labelled
$\{T_a\}$ and $\{\tT_a\}$, respectively ($a=1,2,3$). For convenience
we also split the index $a=(i,3)$, $i=1,2$. The 
commutation relations are given by\footnote{We note that $e_3$ is not
the algebra for the Euclidean group in three dimensions. 
Also, in the rest of the paper 
we will not distinguish between upper and lower Lie-algebra indices.}
\ba
&& [T_a,T_b]=i\e_{abc} T_c ~ , 
\nonumber \\
&& [\tT_3,\tT_i]=\tT_i ~ ,~~~~[\tT_i,\tT_j]=0  ~ .
\label{csu2e}
\ea
The above algebras can be thought of as a pair of maximally isotropic
subalgebras of a 6-dimensional algebra corresponding to a Drinfeld double
Lie group. The complete set of commutation relations is given by \eqn{csu2e}
and the ``mixed'' ones
\ba
&& [T_i,\tT_j] = i \e_{ij} \tT_3 -\d_{ij} T_3~ , ~~~ [T_3,\tT_i] = 
i \e_{ij} \tT_j ~ , 
\nonumber \\
&& [\tT_3,T_i]  = i \e_{ij} \tT_j - T_i~ .
\label{cmix}
\ea
It can be shown that the combined set of commutation relations in \eqn{csu2e}
and \eqn{cmix} obey, the Jacobi identities. 
Next we construct an invariant product for our 6-dimensional Drinfeld double
algebra, with respect to which the various generators should obey \eqn{bili}.
Consider two elements in the Lie algebra of the form $(x_i,y_i)$,
with $i=1,2$, where $x_i \in su(2)$ and $y_i \in e_3$.
Then, the invariant inner product is defined as 
\be
\langle (x_1,y_1)| (x_2,y_2) \rangle = \{x_1,x_2\} - \{y_1,y_2\} ~ ,
\label{indfg}
\ee
where $\{\cdot,\cdot\}$ is 
the invariant product for the algebras $su(2)$ and $e_3$.
A representation for the various generators is given by 
\ba 
&& T_a = ({\s_a\ov 2},{\s_a\ov 2}) ~ , ~~ a=1,2,3 ~ ,
\nonumber \\ 
&& \tT_1 = (\s_+,-\s_-) ~ ,~~~ \tT_2= -i (\s_+,\s_-) ~ ,~~~
\tT_3 = \ha (\s_3,-\s_3) ~ ,
\label{repss}
\ea
where the $\s_a$'s denote the three 
Pauli matrices and $\s_\pm = \ha(\s_1\pm i \s_2)$.
Then, the product $\{\cdot,\cdot\}$ of any two matrices is equal to
their trace.
We note that our double is similar to the $O(2,2)$ non-compact double used in 
\cite{KliSevII}.

Next, we parametrize the $SU(2)$ group element in terms of
the three Euler angles $\phi$, $\psi$ and $\th$. It is represented by 
the $4 \times 4$ block-diagonal matrix 
\be
g_{SU(2)}= \diag( g, g ) ~ ,
\label{gsu2}
\ee
where 
\be
g = e^{{i\ov 2}\phi \s_3}  e^{{i\ov 2}\th \s_2}  e^{{i\ov 2}\psi \s_3} =
\pmatrix{ \cos {\th\ov 2} e^{{i\ov 2} (\phi +\psi)} & 
\sin {\th\ov 2} e^{{i\ov 2} (\phi -\psi)} \cr
- \sin {\th\ov 2} e^{-{i\ov 2} (\phi -\psi)} & 
\cos {\th\ov 2} e^{-{i\ov 2} (\phi +\psi)} \cr } ~ .
\label{su211}
\ee
Also the group element of $E_3$ is parametrized in terms of three 
variables $y_1$, $y_2$ and $\chi$ and represented by the following 
$4\times 4$ block-diagonal matrix 
\be
\tg_{E_3}= \diag(\tg_+,\tg_-) ~ ,
\label{ge3}
\ee
where
\ba
&& \tg_+ = \pmatrix { e^{+ {\chi \over 2}} & \chi_+ \cr
0 & e^{-{\chi\over 2}} \cr} ~ ,~~~~
\tg_-  = \pmatrix { e^{- {\chi \over 2}} & 0 \cr
\chi_- & e^{+{\chi\over 2}} \cr} ~ ,
\nonumber \\
&& \chi_\pm = \pm e^{-{\chi\ov 2}} (y_1 \mp i y_2) ~ .
\label{ge22}
\ea
The Maurer--Cartan forms are defined as $L_a =-i
 \langle g\inv d g| \tT_a \rangle$
and $\tL_a =  \langle \tg\inv d \tg| T_a \rangle$. Using the parametrization
of the $SU(2)$ group element in \eqn{su211} we have explicitly
\ba
L_1 &  =&  \cos \psi \sin \th d\phi - \sin \psi d\th ~ ,
\nonumber \\
L_2 & = &  \sin \psi \sin \th d\phi + \cos \psi d\th  ~ ,
\label{mausu2} \\
L_3 & = & d\psi + \cos \th d\phi ~ .
\nonumber 
\ea
Similarly, using the parametrization \eqn{ge3} for the $E_3$ group element
we find
\be
\tL_1 =   e^{-\chi }  dy_1 ~ ,~~~~ \tL_2  =  e^{-\chi }  dy_2 ~ ,~~~~
\tL_3  =  d\chi ~ .
\label{maue3}
\ee
Also the matrices
\be
(a_{ab}) = \pmatrix { \cos \phi \cos \psi \cos \th - \sin \phi \sin \psi
& \cos \phi \sin \psi \cos \th + \sin \phi \cos \psi &
- \cos \phi \sin \th \cr 
- \sin \phi \cos \psi \cos \th - \cos \phi \sin \psi & 
\cos \phi \cos \psi - \sin \phi \sin \psi \cos \th & \sin\phi \sin\th \cr
\cos\psi \sin \th & \sin\psi \sin \th &  \cos \th \cr } ~ ,
\label{aij}
\ee
and
\be 
(b_{ab}) = \pmatrix { a_{12} + a_{21} & a_{22} - a_{11} & a_{23} \cr
a_{22} - a_{11} & -a_{12} - a_{21} & - a_{13} \cr
a_{32} & -a_{31} & 0 \cr } ~ ,
\label{bij}
\ee
as well as
\be
(\tilde a_{ab}) = \pmatrix { e^{-\chi} & 0 & 0 \cr
0 & e^{-\chi} & 0 \cr
y_1 e^{-\chi} & y_2 e^{-\chi} & 1 \cr }
\label{tiaij} ~ ,
\ee
and
\be
(\tilde b_{ab})= 
\pmatrix { - y_1 y_2 e^{-\chi} & \sinh \chi + {1\ov 2}
( y_1^2 - y_2^2) e^{-\chi} &  - y_2  \cr
- \sinh \chi + {1\ov 2} ( y_1^2 - y_2^2) e^{-\chi} & y_1 y_2 e^{-\chi}
&   y_1 \cr
 y_2 e^{-\chi}  &  - y_1e^{-\chi}  & 0 \cr }~ .
\label{tibij}
\ee
Then the matrices $\Pi = b^T  a $ and  $\tilde \Pi = \tilde b^T \tilde a $ 
can be computed. Since a $3 \times 3$ antisymmetric tensor has as 
many independent components as a three-vector, 
we may represent them in terms of two vectors $\vec A$ and $\vec \tA$ using 
\be
\Pi_{ab} = -\e_{abc} A_c~ ,~~~~~ \tilde \Pi_{ab} = -\e_{abc} \tA_c ~ .
\label{ppip}
\ee
In the case at hand we compute
\ba
\vec A & = & \pmatrix {\cos \psi \sin \th, \sin \psi \sin \th, \cos \th -1}~ ,
\nonumber \\
\vec \tA &  =&  \left( y_1 e^{-\chi}, y_2 e^{-\chi},
\sinh \chi e^{-\chi} -\ha (y_1^2 + y_2^2) e^{-2 \chi}\right )~ .
\label{vecs}
\ea
%
%
%

\subsection{The classical canonical transformation}

Given the underlying algebraic and group theoretical structure, there exists
a classical canonical transformation relating the corresponding $\s$-model
actions \cite{PLsfe1,PLsfe2}. 
This is universal and does not depend on the particular choice of 
the matrix $E_0$ in \eqn{action1}, \eqn{action2}. Moreover it always has the 
same form, even when we include spectator fields.
For $\s$-models related by \PL\ corresponding to three-dimensional groups
it can be shown that the general canonical transformation, given
in \cite{PLsfe1,PLsfe2}, takes the form
\ba
\vec P & = & {1\ov 1-\vec A\cdot \vec \tA} \left( \vec \tA \times \vec L_\s 
+ \vec \tL_\s - (\vec \tA\cdot \vec \tL_\s) \vec A\right)  ~ ,
\nonumber \\
\vec{ \tilde P} & = & {1\ov 1-\vec A\cdot \vec \tA} 
\left( \vec A \times \vec \tL_\s 
+ \vec L_\s - (\vec A\cdot \vec L_\s) \vec \tA \right)  ~ ,
\label{PPP}
\ea
where the three-vectors $\vec P$ and $\vec{\tilde P}$ have components 
given by $P_a= P_\m L\inv_{\m a}$ and 
$\tilde P_a= \tilde P_\m \tL\inv_{\m a }$, with $P_\m$, $\tilde P_\m$ being 
the conjugate momenta to $X^\m$ and $\tilde X^\m$, respectively.
These transformations can be derived from a generating functional of the form
\be
F= \oint d\s ( B_\m \del_\s X^\m 
+ \tilde B_\m \del_\s \tX^\m ) ~ ,
\label{Funcc}
\ee
for some functions $B_\m$, $\tilde B_\m$ of the target space variables, 
which are determined by solving a set of differential equations \cite{PLsfe2}.
For three-dimensional cases these can be cast into the form
\ba 
\del_{[\m} B_{\n]} & = & -\biggl( 1-\vec A\cdot \vec \tA \biggr)\inv 
\vec \tA \cdot \vec L_\m \times \vec L_\n ~ ,
\nonumber \\
 \tilde \del_{[\m} \tilde B_{\n]}&  = & 
\biggl( 1-\vec A\cdot \vec \tA \biggr)\inv 
\vec A \cdot \vec \tL_\m \times \vec \tL_\n ~ ,
\label{diif} \\
\del_\m \tilde B_\n - \tilde \del_\n B_\m &  =& 
\biggl( 1-\vec A\cdot \vec \tA \biggr)\inv 
\left( \vec L_\m\cdot \vec \tL_\n - (\vec A\cdot \vec L_\m) 
(\vec \tA\cdot \vec \tL_\m) \right)~ .
\nonumber
\ea
For the specific models at hand it is possible to find the explicit 
solution to them 
\ba
B_\phi& =& - \ln\left(
e^{2\chi} \cos^2{\th\ov 2} + e^\chi \sin \th (y_1 \cos\psi  
+ y_2 \sin\psi ) + (1+y_1^2+y_2^2) \sin^2{\th\ov 2} \right ) ~ ,
\nonumber \\
B_\psi & = & - \chi + 2 { y_1 \cos\psi + y_2 \sin \psi\ov 
\sqrt{1+(y_1 \sin\psi - y_2 \cos\psi)^2} }
\cot\inv\left( {\sqrt{1+(y_1 \sin\psi - y_2 \cos\psi)^2} \tan{\th\ov 2} \ov 
e^\chi +y_1 \cos\psi + y_2 \sin\psi }\right) ~ ,
\nonumber \\
B_\th & = & 0 ~ ,
\nonumber \\
{\tilde B_{y_1}}& = & 
+ 2 {\sin \psi \ov \sqrt{1+(y_1 \sin\psi - y_2 \cos\psi)^2} }
\tan\inv \left({y_1 \cos\psi + y_2 \sin \psi + e^\chi \cot{\th\ov 2}\ov
\sqrt{1+(y_1 \sin\psi - y_2 \cos\psi)^2}}\right) ~ ,
\nonumber \\
{\tilde B_{y_2}}& = &
- 2 {\cos \psi \ov \sqrt{1+(y_1 \sin\psi - y_2 \cos\psi)^2} }
\tan\inv \left({y_1 \cos\psi + y_2 \sin \psi + e^\chi \cot{\th\ov 2}\ov
\sqrt{1+(y_1 \sin\psi - y_2 \cos\psi)^2}}\right)~ ,
\nonumber \\
{\tilde B_\chi} & = & -\phi ~ .
\label{soll}
\ea
The above generating functional exhibits a generic, for \PL, 
behaviour \cite{PLsfe2}. Namely, it is highly non-linear in the
group variables of $SU(2)$ and $E_3$ and therefore
it will receive quantum corrections when the canonical 
transformation is implemented in the full Hilbert space along the lines
of \cite{qufun}. Nevertheless, its existence seems to guarantee the 
quantum equivalence of the corresponding dually related $\s$-models, as noted
in footnote 2 and as we will see in section 3.

\subsection{Explicit three- and two-dimensional models}

Let us consider three-dimensional dual models with no
spectators at all and choose the constant matrix 
$E_0\inv = \diag(\l_1,\l_2,\l_3)$.
Then the metric and antisymmetric tensor corresponding to the 
action \eqn{action1} can be written as
\ba
ds^2 & = & {1\ov V} \left ( A_a A_b + {\l_1 \l_2 \l_3 \ov \l_a} \d_{ab} \right)
L_a L_b ~ ,
\nonumber\\ 
B & = & {1\ov V} \e_{abc} \l_c A_c L_a \wedge L_b ~ ,
\label{mebv}\\
V & \equiv & \l_1 \l_2 \l_3 + \l_a A_a^2 ~ .
\nonumber
\ea
We note that since there is no explicit dependence of $L_a$ and $A_a$ 
on the Euler angle $\phi$ the background \eqn{mebv} has the corresponding 
isometry. 
For the dual model the expressions for the background metric and antisymmetric
tensor are identical to those in \eqn{mebv}, with tilded
symbols replacing the untilded ones (also $\tilde \l_a = 1/{\l_a}$).
A greater simplification occurs if two of the constant coefficients are
equal, i.e. if $\l_1=\l_2$. It is then easy to see that there is a second 
commuting isometry corresponding to the vector field ${\del \ov \del \psi}$.
Let us reparametrize the constants $\l_1=\l_2$ and $\l_3$ in terms of two 
other constants $\kappa$ and $g$ as $\l_1=\l_2 = \kappa (1+g)^{1/2}$ and
$\l_3 = \kappa (1+ g)^{-1/2}$.
Then the metric and the antisymmetric tensors are given by
\ba
 ds^2 & = &  \kappa\inv (1+ g)^{-1/2} V\inv 
\left ( \kappa^2 (L_a L_a + g L_3 L_3)
+  4 \sin^4 {\th\ov 2} (d\phi -d \psi)^2 \right) ~ ,
\nonumber \\
B&  =&   2 \sin \th  V\inv~ 
d\th \wedge \left( (1-2 g (1+g)\inv \sin^2 {\th\ov 2})  d\phi 
+ d\psi \right)
\label{fgh1} ~ ,
\ea
where
\be
V \equiv \kappa^2 + \sin^2\th + 4 (1+g)\inv \sin^4 {\th\ov 2} ~ ,
\label{vv}
\ee
and 
\be 
L_a L_a + g L_3 L_3 =
(1+g) d \psi^2 + (1+g \cos^2 \th) d \phi^2 + d \th^2 
+ 2 (1+g) \cos \th d \psi d \phi 
\label{fgh}
\ee
is the metric for $S^3$ deformed by the bilinear in the current $L_3$.
Notice that the antisymmetric tensor is a pure gauge,
i.e. $dB=0$.
Also the dual metric and the dual antisymmetric tensors are given by
\ba
d\tilde s^2 & = & \kappa^3 (1+g)^{1/2} \tilde V\inv
\biggl\{ e^{-2 \chi} \left(e^{- \chi} (y_1 dy_1 + y_2 dy_2 ) 
 + \left(\sinh \chi -\ha 
(y_1^2 + y_2^2 ) e^{-\chi} \right) d \chi \right)^2 
\nonumber\\
&&
{}\qq\qq\qq~~~ + \kappa^{-2} (1+g)\inv d\chi^2 + 
\kappa^{-2}  e^{-2 \chi} ( dy_1^2 + d y_2^2) \biggr\} ~ ,
\nonumber \\
\tilde B & = & 2 \kappa^2 e^{-2 \chi} \tilde V\inv 
\biggl\{ (1+g) e^{-\chi} \left(\sinh \chi 
-\ha (y_1^2 + y_2^2) e^{-\chi}\right ) dy_1 \wedge dy_2
\nonumber \\
&& {}\qq\qq~~~~  + (y_1 dy_2 -  y_2 dy_1) \wedge d\chi \biggr\}~ ,
\label{dsssa}
\ea
where 
\be \tilde V \equiv   1 + \kappa^2 e^{-2 \chi} \left(y_1^2 + y_2^2
+ (1+ g) \left( \sinh \chi - \ha (y_1^2 + y_2^2) e^{-\chi}\right)^2
\right) ~ .
\label{vvv}
\ee

Notice that if we take the limit $\kappa \to \infty$ and also rescale the
overall coupling constant as $\l\to \l \kappa\inv (1+g)^{-1/2}$, 
the metric in \eqn{fgh1} reduces to the deformed $S^3$ metric in \eqn{fgh}.
In addition, we rescale $(y_1,y_2,\chi)\to {1\ov \kappa} (x_1,x_2,x_3)$. This 
contracts the group $E_3$ into an  Abelian one.
Then, \eqn{dsssa} 
reduces to the expression for the usual non-Abelian dual of \eqn{fgh} with
respect to the left action of $SU(2)$. 
If on the other hand we rescale as
$\th \to \kappa \r$, $\phi \to \ha(\kappa x_3 + 2 \a)$,
$\psi \to \ha(\kappa x_3 - 2 \a)$
and $\l\to \kappa  \l (1+g)^{1/2}$,
and if we let then $\kappa \to  0$ we obtain the non-Abelian dual of $E_3$ 
(effectively this contracts also $SU(2)$ into an Abelian group). 
In addition, for both of the limits we just briefly mentioned, the generating 
functional \eqn{soll} reduces to the appropriate one for non-Abelian duality
\cite{zacloz,loz}.

We may also consider a different limit. Namely we reparametrize 
$\kappa = 2 a\inv (1+g)^{-1/2} $ and then let $g\to -1$. This implies that 
$\l_1 = \l_2 = 2/a$ are finite in that limit, whereas 
$\l_3 = 2 a\inv (1+g)\inv $ is sent to infinity. We also rescale the overall 
coupling constant as $\l\to \l a/2$. The resulting models have 
two-dimensional target spaces as the third dimension becomes suppressed in 
that limit. The metric corresponding to \eqn{fgh1} becomes 
\be
ds^2 = {1\ov 1 + a^2 \sin^4{\th\ov 2}} \left(d \th^2 
+ \sin^2\th d\phi^2\right)~ .
\label{ghfa}
\ee
It represents a deformed 2-sphere, as can also be
seen by explicitly verifying that the Euler characteristic $\chi=
{1\ov 4 \pi} \int \sqrt{G}R $ equals 2.
The metric corresponding to \eqn{dsssa} is given by 
\be 
d\tilde s^2 = {1/2 \ov r ( 1+ a z) } \left( dz^2 + \biggl( dr 
+ {z- a r/2\ov 1+ a z} dz \biggr)^2 \right) ~ ,
\label{jw}
\ee
where
we have changed variables as $y_1^2 + y_2^2 = \ha r a^2 $ 
and $e^{2\chi}= 1+ a z$. This is a non-compact manifold.
In the limit $a\to 0$, \eqn{ghfa} and \eqn{jw} reduce to the metric 
for $S^2$ and its non-Abelian dual with respect to $SU(2)$ (see third 
article in ref. \cite{nonabel}).

In order to perform the 1-loop quantum analysis in a unified manner 
it will be useful to cast the metrics \eqn{ghfa} and \eqn{jw}
into the conformally flat form
\be 
ds^2 = e^{-2 \Phi} (dx^2 + dy^2 ) ~ ,
\label{fhv}
\ee
where the appropriate change of variables and the inverse of
the conformal factor are given for \eqn{ghfa} by
\be
x =  \cos\phi \cot{\th \ov 2}~ , ~~~~~ y~ =~ \sin\phi \cot{\th\ov 2}  ~ ,
\label{ch11}
\ee
and 
\be
e^{2\Phi} =  {1\ov 4}\left( (1 + x^2 + y^2 )^2 + a^2\right) ~ ,
\label{ch1}
\ee
and for \eqn{jw} by 
\ba 
x & = & {1\ov a^2}\left( (1+a z)^{1/2} -2 \right) +
\left ({1\ov a^2} + {r\ov 2}\right ) ( 1+ a z )^{-1/2} ~ ,
\nonumber \\
y & = & {1\ov a}\left( (1+a z)^{1/2} -1 \right) ~ ,
\label{ch22} 
\ea
and 
\be
e^{2 \Phi}  =  x (1+ a y) - y^2~ .
\label{ch2}
\ee
%
The two-dimensional
models \eqn{ghfa} and \eqn{jw} correspond to some analytic continuations
of ``dressed coset'' models in \cite{KliSevdressed}.\footnote{A
comparison of \eqn{jw} with the corresponding metric (after analytic 
continuation) in \cite{KliSevdressed} is facilitated
when the latter is also cast into the form \eqn{fhv}. The conformal factor 
is exactly the same as that in \eqn{ch2}, but the necessary 
coordinate transformation to bring the metric into that form is different
than \eqn{ch22}.}

\section{Renormalization}

In this section we investigate the behaviour under the renormalization
group of our 2-dimensional dually related models \eqn{fhv} with conformal
factors given by \eqn{ch1} and \eqn{ch2}. For the cases of the usual Abelian
and non-Abelian dualities, 
a similar investigation was performed for a one-parameter family of
deformations
of the Principal Chiral model for $SU(2)$, with metric given in \eqn{fgh},
and its non-Abelian dual \cite{zacloz,ouggroi} in \cite{ouggroi}.

Consider a two-dimensional $\s$-model with Lagrangian density 
$\cL = {1\ov 2 \l} Q^+_{\m\n} \del_+ X^\m \del_- X^\n$, where $Q_{\m\n}^+=
G_{\m\n}+B_{\m\n}$. It will be renormalizable
if the corresponding counter-terms, at
a given order in a loop expansion, 
can be absorbed into a renormalization of the 
coupling constant $\l$ and (or) of 
some parameters labelled collectively $a$. 
In addition, we allow for general field redefinitions, which are 
coordinate reparametrizations in the target space.
This definition of renormalizability of $\s$-models is quiet strict and
similar to that for ordinary field theories. A natural extension of this is to
allow for the manifold to vary with the mass scale and the renormalization 
group to act in the infinite dimensional space of all metrics and torsions
\cite{Frialv}. Further discussion of this generalized renormalizability will
not be needed for our puproses.
Perturbatively in powers of $\l$ we express the bare quantities, denoted 
by a zero as a subscript, as
\ba
&&\l_0 = \m^\e \l \left(1+ {J_1(a)\ov \pi \e} \l + {J_2(a)\ov 8 \pi^2 \e} \l^2
\cdots \right) \equiv \m^\e \l
\left (1+ {y_\l \ov \e} + \cdots \right )  ~ ,
\nonumber \\
&&a_0 = a + {a_1(a) \ov \pi \e} \l  + {a_2(a)\ov 8 \pi^2 \e} \l^2
+ \cdots  \equiv a 
\left (1+ {y_a \ov \e} + \cdots \right )  ~ ,
\label{expp}\\
&&X^\m_0 = X^\m + {X^\m_1(X,a) \ov \pi \e} \l +  {X^\m_2(X,a)\ov 8 \pi^2 \e} 
\l^2+ \cdots  ~ .
\nonumber
\ea
The ellipses stand for higher-order loop- and pole-terms in $\l$ and
$\e$ respectively.
Then, the beta-functions up to two loops are given by 
$\b_\l = \l^2 {\del y_\l\ov \del \l}
= {\l^2\ov \pi}\left(J_1 + J_2 \l/(4\pi)\right)$ and 
$ \b_a = \l a {\del y_a\ov \del \l}=
{\l \ov \pi} \left(a_1 +a_2 \l/(4\pi)\right) $.
The equations to be satisfied by appropriately choosing $J_i,a_i$ and $X_i^\m$,
with $i=1,2$, are given by
\be
T^{(i)}_{\m\n}=
-J_i Q^+_{\m\n} + \del_{a} Q^+_{\m\n} a_i + \del_\l Q^+_{\m\n} X_i^\l
+  Q^+_{\l\n} \del_\m  X_i^\l  +  Q^+_{\m\l} \del_\n X_i^\l~ , ~~~~~~ i=1,2~ ,
\label{1loop}
\ee
where the corresponding counter-terms computed in the dimensional
regularization \break scheme are (see, for instance, \cite{Osborn})
\ba
&& T^{(1)}_{\m\n} = \ha R^-_{\m\n} ~ , ~~~~
T^{(2)}_{\m\n}= {1\ov 4} R^-_{\m\l\r\s} Y^{\r\s\l}{}_\n~ , 
\nonumber \\
&& 
Y_{\r\s\l\n} \equiv -2 R^-_{\r\s\l\n} + 3 R^-_{[\l\r\s]\n} + \ha (H^2_{\l\r}
G_{\s\n} -H^2_{\l\s} G_{\r\n})~ ,
\label{effg}
\ea
with $R^-_{\m\n\r\l}$ and $R^-_{\m\n}$ being the ``generalized'' curvature and 
Ricci tensors constructed with connections that include the torsion.
In principle we may add a term on the right-hand side corresponding to a
redefinition (gauge transformation)
of the antisymmetric tensor as $B_{\m\n}\to B_{\m\n} + \del_{[\m} \L_{\n]}$.
Such a gauge transformation will affect only the antisymmetric part of 
\eqn{1loop}.
 
Next we specialize to the case of two-dimensional metrics 
of the form \eqn{fhv} with ($X^1=z=x+i y$, $X^2=\z=x-i y$).
If we work out \eqn{1loop} for $\m=\n = 1$ and 
$\m=\n=2$, we obtain the conditions $\del_z \z_i= 0 $ and $\del_\z z_i = 0$
(for $i=1,2$) respectively. Setting $\m=1$, $\n =2$ we obtain instead
\ba
e^{2\phi} \del_{z}\del_{\z} \Phi & = &
 - {J_1\ov 2} - \del_{a} \Phi a_1 
+ \left( \ha \del_z - \del_z\Phi\right) z_1 
+ \left( \ha \del_\z - \del_\z\Phi\right) \z_1~ ,
\label{conn1}\\
8 e^{4\phi} (\del_{z}\del_{\z} \Phi)^2 & = &
 - {J_2\ov 2} - \del_{a} \Phi a_2 
+ \left( \ha \del_z - \del_z\Phi\right) z_2 
+ \left( \ha \del_\z - \del_\z\Phi \right) \z_2~ .
\label{conn2}
\ea
For two-dimensional models the antisymmetric tensor is a pure gauge. It has no
effect on \eqn{1loop} since the latter has only a symmetric part.
%
At the 1-loop level, there remains to compute the 
constants $J_1,\e_1$ and the holomorphic 
function $z_1(z)$. The function $\z_1(\z)$ is the complex conjugate 
of $z_1(z)$.
For the model \eqn{ghfa} we found, using \eqn{conn1}, that
\ba
&& J_1 = -\ha (1+a^2)~ , ~~~~~ a_1 = \ha a (1+a^2) ~ ,
\nonumber \\
&&  x_1 = {1\ov 4} a^2 x~ , ~~~~~ y_1 = {1\ov 4} a^2 y~ ,
\label{ress1}
\ea 
where $z_1 = x_1 + i y_1$.
The corresponding computation for the model \eqn{ch2} gives
\ba
&& J_1 = -\ha (1+a^2)~ , ~~~~~ a_1 = \ha a (1+a^2) ~ ,
\nonumber \\
&&  x_1 = {1\ov 4} -\left(1+ {3\ov 4} a^2\right) x -{a\ov 4} y  ~ , ~~~~~ 
y_1 = - {a \ov 4} +{a\ov 4} x -\left(1+ {3\ov 4} a^2\right) y~ .
\label{ress2}
\ea 
Since $J_1$, $a_1$ are the same for both models, we conclude that they have 
the same 1-loop beta-functions, which are explicitly given by\footnote{We 
have also checked the renormalizability of the models at the 2-loop level 
and found that \eqn{conn2} admits no solution
with the exception of the model \eqn{ghfa},
which becomes renormalizable when $a=0$ (with $J_2= -1$, in agreement with 
\cite{ouggroi}). 
Hence, the corresponding counter-term cannot be absorbed into a
renormalization of the couplings and of the fields.
This presumably implies that the
corresponding \PL\ transformation rules 
need to be corrected order by order in perturbation theory as it is the 
case for ordinary T-dualities \cite{tse,ouggroi}.
Such a conclusion is also supported by the expectation 
that the non-linear generating  
functional \eqn{soll} will receive quantum corrections \cite{qufun}.
This implies that the induced canonical transformation, hence the \PL\
rules, will be accordingly corrected.}
\be
\b_\l = - {\l^2 \ov 2\pi} (1+a^2)~ , ~~~~~
\b_a = {\l \ov 2 \pi} a ( 1+ a^2)~ .
\label{1kjd}
\ee
Notice that according to this system of equations $a$ runs to infinity and 
the product $\l a $ remains constant. Hence, under renormalization flow,
$\l$ approaches zero. However, in the ultraviolet the fixed point is not a
trivial one. This is easily seen by performing 
the scaling of variables $x\to x a^{1/2}$ and $y\to y a^{1/2}$ in
\eqn{ch1} and \eqn{ch2}, and then taking the limit $a\to \infty$. 
We also have to rescale the 
coupling $\l\to \l/a$, so that the product $\l a$ is kept fixed. In the limit
$a\to \infty$ the resulting metrics are well defined and 
again of the form \eqn{fhv}, with the
conformal factors given by $e^{2\Phi_{\infty}}=1/4(1+(x^2 +y^2)^2)$
(also with the topology of the 2-sphere) and
$\e^{2\Phi_{\infty}}= x y$ respectively. Since these do not correspond to
flat metrics, the theories do not become asymptotically free.
This is to be contrasted with the behaviour of $\b_\l$ for the $O(3)$-chiral 
model and its non-Abelian dual that is obtained from $\b_\l$ in \eqn{1kjd}
for $a=0$. For these theories the coupling runs to zero in the ultraviolet
and therefore they 
are asymptotically free. Hence, at the quantum level, a limit 
under which \PL\ reduces to the usual non-Abelian duality does not exist.
We also note that the $\s$-models corresponding to the geometries 
with the $\Phi_\infty$'s above
are no longer renormalizable. This is natural as they correspond 
to the fixed point at the ultraviolet of the geometries corresponding to
the $\Phi$'s, but it can also be checked independently by seeing that 
there are no $J_1$, $a_1$ and $z_1$ that solve \eqn{conn1}.

We have mentioned
that the Drinfeld double we have considered, in the limit of $SU(2)$
contracting into an Abelian group, will be appropriate for $\s$-models 
related by non-Abelian duality with respect to the $E_3$ group.
As was shown in \cite{nonsemi}, such models are problematic 
as far as conformal invariance is concerned, since the 
algebra structure constants are not traceless. This is not in conflict with 
our result that for a particular ``non-conformal'' example 
the 1-loop beta-functions are
equivalent. As we have seen, taking singular limits does not always produce 
quantum mechanically the expected result from classical 
considerations.\footnote{Nevertheless, this was the
case with the singular limits considered in \cite{planew,gausfe}.
In these cases one starts with exact string solutions and, after the 
limit is taken, one obtains plane wave (among other)
exact solutions to string theory as well. Hence, in these examples conformal
invariance was already an input before any limits were taken.}
Contraction of $SU(2)$ into an  
Abelian group is certainly such a limit. This issue does deserve 
further investigation.

There are several directions in which the present work can be extended.
One should further investigate the conditions \eqn{1loop} in full 
generality and classify all $\s$-models related by \PL\ that have equivalent
beta-functions, as in the example we have considered. An important question
is whether or not the limit of \PL\ to non-Abelian duality,
i.e. when one of the groups becomes Abelian, exists quantum mechanically.
Our present investigation suggests that this is not always the case,
but one would like to know to 
what extent this is a general statement and not a model-dependent one.
Finally, the construction of conformal $\s$-models related 
by genuine \PL\ is an important open problem since
it is of direct relevance to 
string theory. An important step towards this will be
the complete classification of all Drinfeld doubles.

\bs\bs
\centerline{\bf Acknowledgements}

I would like to thank L. Alvarez-Gaum\'e and C. Kounnas 
for discussions, and L. Palla and A.A. Tseytlin for e-mail correspondence.
Three-dimensional $\s$-models based on the Drinfeld double of subsection 2.1
have been also constructed in \cite{toappp}, where
the interplay between \PL\ and the renormalization group 
has been also considered.

\bs\bs
\centerline{\bf Note added }

1. The comments of this paragraph follow a discussion with J. Iliopoulos.

The fact that the renormalization group flow drives the $\s$-model
\eqn{ghfa} (and also \eqn{jw}) away from the point $a=0$
is quite similar to phenomena discussed some years 
ago in the context of usual four dimensional field theories, in searching 
for the origin of symmetries in Nature (see \cite{Iliop} and refs. therein). 
In some of these examples,
enhanced symmetry points in the space of couplings and parameters acted as 
infrared repulsors due to the fact that the
theories corresponding to the enhanced symmetry points were asymptotically 
free.
In our case the enhanced symmetry point, with an $SU(2)$ symmetry group,
is at $a=0$ and the corresponding
$\s$-model for the $S^2$ metric is indeed asymptotically free.

2. The referee suggested that the reason that the $\s$-models \eqn{ghfa} and 
\eqn{jw} are not 2-loop renormalizable, in the strict field theoretical sense,
might be due to the fact that we have not allowed for the most general 
renormalization of constants possible. 
Indeed, there is a more general class of 
2-dimensional $\s$-models that are related 
by \PL\ (see also \cite{KliSevdressed}).
They can be obtained by
taking a non-diagonal matrix $E_0$ in \eqn{action1} and \eqn{action2}. 
For our purposes it is enough to concentrate on the
generalization of \eqn{ghfa}, which reads 
$$ds^2 ={1\over 1+ a^2 (b-\sin^2{\th\over 2})^2 }
 \left(d\th^2 + \sin^2\th d\phi^2\right)\ ,$$
where $a,b$ are constants. It is easy to check that this is the most 
general form preserving the isometry corresponding to shifts of 
$\phi$ and also keeping $e^{2 \Phi}$ a quadratic in 
$z\z$ polynomial, when the metric is written in the form \eqn{fhv} after 
using \eqn{ch11}.
It is a lengthly, but straightforward, calculation to verify that
the condition for a 2-loop renormalizability \eqn{conn2}
is not satisfied for any choice of $a_2,b_2,J_2$ and $z_2,\z_2$.

Hence, the only reasonable conclusion is that the \PL\ transformations have 
to be modified to this order. 
This is in agreement with the general arguments presented in footnote 6.



\end{document}